\documentclass[twocolumn]{svjour3}

\usepackage[T1]{fontenc}
\usepackage[utf8]{inputenc}
\usepackage{graphicx} 
\usepackage{bbding} 
\usepackage[english]{babel}
\usepackage{multicol}
\usepackage{framed}
\usepackage{multirow}
\usepackage{listings}
\usepackage[hyphens]{url}
\usepackage[]{hyperref}
\hypersetup{
  colorlinks = true,
    linkcolor = black,
    citecolor = black,
    filecolor = black,
    urlcolor = black,
}
\usepackage{tikz} 
\usetikzlibrary{decorations.text}
\usetikzlibrary{patterns}
\usetikzlibrary{calc,trees,positioning,arrows,chains,shapes.geometric,%
    decorations.pathreplacing,decorations.pathmorphing,shapes,%
    matrix,shapes.symbols}
\usepackage{amsmath,bm,times}
\usepackage{framed}
\usepackage{bbding}
\usepackage{amsmath}

\usepackage[sorting=none,firstinits=true,doi=false,url=false, isbn=false]{biblatex}
\bibliography{bib/bib.bib}

\newcounter{observations}
\newcommand{\observation}{\stepcounter{observations} \vspace{.3cm}\noindent{\bf Observation \arabic{observations}} }
\newcounter{conclusions}
\newcommand{\conclusion}{\stepcounter{conclusions} \vspace{.3cm}\noindent{\bf Conclusion \arabic{conclusions}} }

\newcommand{\spo}{\emph{smartphone1}}
\newcommand{\spt}{\emph{smartphone2}}
\newcommand{\tabone}{\emph{tablet1}}

\newcommand{\ucad}[0]{Ad\-Remover}
\newcommand{\ucperm}[0]{BetterPermissions}

\newcommand{\appName}[0]{NewsReader}
\newcommand{\instrumentedAppName}[0]{Instrumented \appName{}}

\begin{document} 


\title{In-Vivo Bytecode Instrumentation for Improving Privacy on Android Smartphones in Uncertain Environments}


\author{Alexandre Bartel \and 
Kevin Allix \and
Martin Monperrus \and
Jacques Klein \and
Yves Le Traon} 

\institute{Alexandre Bartel, Kevin Allix, Jacques Klein and Yves Le Traon \at
       {University of Luxembourg, SnT}\\
       {Luxembourg, Luxembourg}\\
       \email{firstName.lastName@uni.lu}
\and Martin Monperrus\at 
       {University of Lille}\\
       {INRIA}\\
       {Lille, France}\\
       \email{martin.monperrus@univ-lille1.fr}
}
\date{Received: date / Accepted: date}


\maketitle
\begin{abstract}
In this paper we claim that an efficient and readily applicable means to improve privacy of Android applications is:
1) to perform runtime monitoring by instrumenting the application bytecode and 
2) in-vivo, i.e. directly on the smartphone.
We present a tool chain to do this and present experimental results showing that this tool chain can run on smartphones in a reasonable amount of time and with a realistic effort.
Our findings also identify challenges to be  addressed before running powerful runtime monitoring and instrumentations directly on smartphones.
We implemented two use-cases leveraging the tool chain: \ucperm{}, a fine-grained user centric permission policy system and \ucad{} an advertisement remover.
Both prototypes improve the privacy of Android systems thanks to in-vivo bytecode instrumentation.\\
\mbox{}
\end{abstract}



\section{Introduction} 
\label{sec:intro}

Android is one of the most widespread mobile operating system in the world accounting for more than 72\% of the market share \cite{gartnerNov2012}. 
More than 500\,000 Android applications available on dozens of application markets can be installed by end users. 
On the official market of Google (Google Play, formerly AndroidMarket), more than 10\,000 new applications are available every month.\footnote{http://www.appbrain.com/stats/number-of-android-apps} 
For the end user, downloading an application on her smartphone is similar to choosing an apple on an apple tree: she only sees the surface and has no evidence that there is no worm in it.   
Unfortunately there are many worms of different kinds waiting to infect smartphones such as malware leaking private data and adware calling premium-rate numbers. 

In this paper we claim that an efficient and readily applicable means to improve privacy of Android applications is to perform runtime monitoring and interception of the application interactions with the Android stack by instrumenting the application bytecode directly on the smartphone (in-vivo).
Before further introducing our contribution let us defend our key claim. 
 
\emph{Why performing runtime monitoring and interception?}
We want to allow or disallow behaviors of an application at runtime.
We use runtime monitoring as it consists of observing the behavior of an application during execution.
It collects certain metrics or intercepts all exchanges at the interface between the application and the rest of the system. 
In this paper, we discuss two case-studies involving runtime monitoring and interception, including an implementation of a fine-grained permission model on top of the Android software stack as proposed in \cite{Computer2012_Stavrou}.

\emph{Why performing bytecode instrumentation?}
There are at least two ways to perform runtime monitoring and interception:
modification of the Android software stack or bytecode instrumentation.
Modification of the software execution stack consists in altering the operating system or the core libraries to intercept the required information.
On Android, it means changing the underlying kernel, the Dalvik virtual machine or the Android framework. 
Unless convincing the Android consortium, this is rather limited in deployment since normal end-users have neither the rights (jailed phones) nor the ability to do so. 
Also, this solution would require users to change their firmware which is a non-trivial task, further complicated by the so called \emph{fragmentation problem} of the Android system as there is not a single Android system but many different Android systems each customized to run on a specific device (tablet, smartphone,~\ldots).
If the operating system is modified, one would need to create a custom instrumented version for every possible Android version which is not easily doable in practice.
Bytecode instrumentation however, is one of the lightest way to perform runtime monitoring on top of execution platform that can not be modified.
In the context of a fine-grained policy enforcement for improving privacy, we are able -- thanks to bytecode instrumentation -- to enforce a fine-grained permission model of already deployed applications on Android smartphones without any modification of the Android software stack.

\emph{Why performing in-vivo instrumentation directly on smartphones?}
Bytecode instrumentation could be done outside the device for instance using a remote service on the Internet.
However, many countries forbid distributing binaries to third-party services (e.g. France).
Also, terms of service of several markets (e.g. Google Play for Android) do not allow this.
Instrumenting applications directly on the device keeps the application within the device.

To sum up, we believe that the most efficient and practical way for ensuring security and privacy on mobile devices is to instrument the application bytecode directly on the smartphone (in-vivo), the instrumentation being tailored for a given security or privacy concern.
Our main contributions are that:
\begin{itemize}
  \item We have built a toolchain to automatically repackage Android applications directly on an Android device;
  \item We have built a toolchain to automatically analyze Android applications directly on an Android device;
  \item The toolchain has been tested by implementing two prototypes which increase the end-user privacy. One removes advertisement and the other gives the user total control over the applications' runtime permissions.
  \item The feasibility of such a tool chain has been evaluated. Limitations and challenges have been pinpointed.
\end{itemize}

To the best of our knowledge, we were the first\footnote{we published a technical report in May 2012 \cite{bartel2012invivo}} to present a tool chain to automatically transform Android applications directly on a device.

The paper is organized as follows:
Section \ref{sec:useCases} provides the reader with two scenarios motivating the need of bytecode instrumentation of Android applications. 
Section \ref{sec:methodology} describes a tool chain for instrumenting Android applications directly on Android devices (smartphones, tablets, ...).
Section \ref{sec:useCases-implem} presents the design and implementation of valuable bytecode instrumentations for the security and privacy of smartphones.
Section \ref{sec:eval} demonstrates the feasibility of running the whole tool chain in a reasonable amount of time.
Section \ref{sec:related-work} discusses the related work and Section \ref{sec:conclusion} concludes the paper.


\section{Use Cases of In-Vivo Instrumentation}
\label{sec:useCases}

There are different scenarios in which it would be beneficial to manipulate and analyze Android applications' bytecode directly on smartphone devices (in vivo). 
In this Section we present two valuable use cases: \emph{AdRemover} and \emph{\ucperm{}}.

Both of them improve the privacy for the user. 
\emph{AdRemover} hinders advertisement libraries to work and thus, at the same time, prevents them from sending private information related to localization (GPS coordinates,...) or of the device itself such as the IMEI (International Mobile Equipment Identity). 
\emph{\ucperm{}} gives users the power to enable or disable applications' permissions.
In an extreme case where the user would like no application to have access to her contact list, she would remove the contact permission from all applications on the phone.
The result is a better privacy for the user.


\subsection{\label{sec:adremove}Advertisement Removal}
\newcommand{\kwarn}[1]{\textcolor{orange}{#1}}

Nearly half of the Android applications embeds third-party code to handle in-app advertisement \cite{addroid}. 
A significant proportion of ad-supported apps include at least two advertising libraries \cite{adsplit}. 

Furthermore, Android applications are distributed as self-sufficient packages, bundling together both specifically developed code and the third-party libraries they may need, such as binary-only advertisement modules. 

Android enforces a per-application policy-based security model: either all parts of an application benefit from a given permission, or none of its parts. 
It means that when a user grants permissions to an application, she actually grants permissions to components potentially written by different entities, including the ad libraries.
 
For example, a newspaper app may be allowed to send its location back to the app publisher so that she is presented with local news.
However, from a privacy perspective the embedded advertisement library should not be allowed to send the location data to the ad companies. 
Currently, the user faces a dilemma: she either has to reduce her privacy level expectation, or refrain from using an otherwise valuable application. 

A workaround of this limitation of the platform is to disable the use of the ad library in-vivo.

This may have positive side-effects, since advertisement libraries also have a significant impact on the battery usage.
According to a recent study~\cite{Pathak2012}, third-party advertisement modules can be held responsible for up to 65\%-75\% of energy spent in free 
applications .


\subsection{Fine-Grained Permission Policy}

The Android framework relies on a permission-based model and follows an \emph{``all or nothing''} policy. 
At installation time, users must either accept or reject all permissions requested by the application. 
An application is installed only if all the requested permissions are accepted.
There is no way to accept only some permissions (such as accessing the localization data) and not others (such as connecting to the Internet).
Users are doomed to completely trust the application developers who write the list of permission. 
Enck et al. \cite{Enck2009a} have pointed out that an application with several sensitive permissions is a real security threat. 
For instance if an application requests the permission to send SMS and a permission to read the contact list, the contact list could potentially be sent to a remote phone by sending it through SMS.

A fine-grained permission model consists in giving users the ability to specify their own set of permissions to applications, according to their own usage.
All sets of permissions for all applications on the device constitutes the security policy.
The underlying permission-based system would then enforce this user-defined policy.

Running such user-level security policy is impossible on a unmodified Android platform with unmodified application code.
However, as we show later, it is indeed possible by manipulating the application bytecode at installation time, in-vivo.




\section{Toolchain for In-vivo Bytecode Instrumentation}
\label{sec:methodology}

\begin{figure}
\begin{center}
\resizebox{\columnwidth}{!}{

\newcommand{\UCFPSOFFSET}{5.4}
\newcommand{\POLICYOFFSET}{5.6}

\pgfdeclarelayer{background}
\pgfdeclarelayer{foreground}
\pgfsetlayers{background,main,foreground}
\tikzstyle{block} = [draw,fill=white!20,minimum width=4em]
\tikzstyle{steps} = [circle, draw, above, inner sep=1pt, midway, yshift=.2cm, fill=black!50, text=white]
\begin{tikzpicture} [node distance=.5cm, start chain=going below, remember picture]

		  \tikzstyle{vertex}=[circle,fill=black!25,minimum size=14pt,inner sep=0pt]
		  \tikzstyle{vInterface}=[block,fill=black!25,minimum size=14pt,inner sep=0pt]
		  \tikzstyle{vPermission}=[diamond,fill=blue!25,minimum size=14pt,inner sep=0pt]
		  \tikzstyle{bytecode}=[]
  		\tikzstyle{tuborg}=[decorate]
  		\tikzstyle{tubnode}=[midway, right=2pt]
			\tikzstyle{stub} = [rectangle, fill=yellow!5, text width=1.8cm, minimum height=.5cm, node distance=.6cm, draw];
			\tikzstyle{filebox} = [rectangle, text width=1.8cm, draw, minimum height=2.3cm, node distance=4cm];
			\tikzstyle{insideBox} = [rectangle, fill=white, text width=1.6cm, minimum height=.3cm, draw, node distance=.2cm];

		\node[filebox, label=above:(a) Original Apk] (Box-originalApk) at (0,0) {
			\tikz \node[insideBox] (nClassesDex){classes.dex};
			\tikz \node[insideBox] {{\tiny AndroidManifest.xml}};
			\tikz \node[insideBox, minimum height=.8cm] {Data};
			\tikz \node[insideBox] {Signatures};
		};

		\node[filebox, right of=Box-originalApk, label=above:(b) Jar File] (Box-jarFile) {
			\tikz \node[insideBox] {Class1.class};
			\tikz \node[insideBox] {Class2.class};
			\tikz \node[insideBox] {Class3.class};
			\tikz \node[insideBox, draw=white] {$\hdots$};
			\tikz \node[insideBox] {ClassN.class};
		};

		\node[filebox, right of=Box-jarFile, label=above:(c) Modified Jar File] (Box-modifiedJarFile) {
			\tikz \node[insideBox] {Class1.class$^*$};
			\tikz \node[insideBox] {Class2.class};
			\tikz \node[insideBox] {Class3.class$^*$};
			\tikz \node[insideBox, draw=white] {$\hdots$};
			\tikz \node[insideBox] {ClassN.class};
		};

		\node[filebox, below = .4cm of Box-modifiedJarFile, label=below:(d) New Dex File] (Box-newDexFile) {
			\tikz \node[insideBox] {Class1$^*$};
			\tikz \node[insideBox] {Class2};
			\tikz \node[insideBox] {Class3$^*$};
			\tikz \node[insideBox, draw=white] {$\hdots$};
			\tikz \node[insideBox] {ClassN};
		};

		\node[filebox, left of=Box-newDexFile, label=below:(e) New Apk File] (Box-newApkFile) {
			\tikz \node[insideBox] (nNewClassesDex) {classes.dex$^*$};
			\tikz \node[insideBox] {{\tiny AndroidManifest.xml}};
			\tikz \node[insideBox, minimum height=.8cm] {Data};
			\tikz \node[insideBox, draw=transparent!100] {\phantom{Signatures}};
		};

		\node[filebox, left of=Box-newApkFile, label=below:(f) New Signed Apk] (Box-newSignedApkFile) {
			\tikz \node[insideBox] {classes.dex$^*$};
			\tikz \node[insideBox] {{\tiny AndroidManifest.xml}};
			\tikz \node[insideBox, minimum height=.8cm] {Data};
			\tikz \node[insideBox] {Signatures$^*$};
		};

\begin{pgfonlayer}{background}
\end{pgfonlayer}
	\draw[-latex] (Box-originalApk) --  node[yshift=-.2cm] {dex2jar} (Box-jarFile);
	\draw[-latex] (Box-jarFile) -- node[yshift=.2cm] {Soot} node[yshift=-.2cm] {ASM} (Box-modifiedJarFile);
	\draw[-latex] (Box-modifiedJarFile.east) -- +(.5cm,0cm) node[xshift=.3cm, yshift=-1.5cm] {dx}|-   (Box-newDexFile.east);
	\draw[-latex] (Box-newDexFile) -- node[yshift=-.2cm] {custom zip}  (Box-newApkFile);
	\draw[-latex] (Box-newApkFile) -- node[yshift=.2cm] {custom} node[yshift=-.5cm, xshift=.13cm, text width=1.3cm] {-keytool\\ -jarsigner} (Box-newSignedApkFile);
\end{tikzpicture}
}
\caption{\label{fig:generateNewApk}Our Process to Instrument Android Applications}
\end{center}
\end{figure}
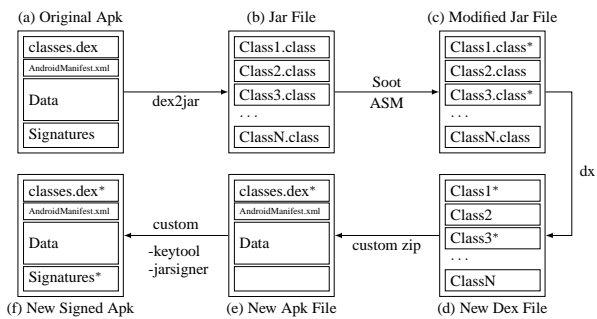

This section presents our proposal for performing bytecode instrumentation of Android applications in vivo, i.e. directly on smartphones.

\subsection{Android Apps in a Nutshell}
\label{sec:android-in-brief}
Android applications are written in Java, compiled into Java bytecode and finally converted to Dalvik bytecode, a bytecode format optimized for embedded devices.
An Android application is a signed zip file (called apk or AndroidPacKage file) containing the Dalvik executable, the \texttt{Android\nolinebreak[0]Manifest\nolinebreak[0].xml} (application metadata), data (e.g. images), and the public key needed to check the provided signatures of all files.

Applications are made of four different components (software abstractions): Activity for the user interfaces, Service for background operations, Content Provider to save data and Broadcast Receiver to receive messages from the system.
The bytecode of an Android application interacts with the Android system through the Android framework (also referred to as the Android API).
Some methods of the API are protected by permissions because they access system resources such as GPS coordinates or the contact list.
Only an application with the right permission(s) can access a protected resource.

\subsection{Requirements}
\label{reqs}

Instrumenting and repackaging a fully-runnable Android application is not straightforward.
It consists of extracting the executable code from the application code, analyzing and instrumenting it, rebuilding a new working android application and signing it again, since the OS requires applications to be signed.

Our toolchain has the following requirements:

\begin{enumerate}\itemsep0em
\item The Android OS must be unmodified (for the sake of a broad applicability as presented in Section \ref{sec:intro});
\item The Dalvik virtual machine that runs Android applications must be unmodified, in particular in terms of configuration values such as the maximum heap size (for the sake of a broad applicability, see Section \ref{sec:intro});
\item The hardware that is used to instrument bytecode must be representative of common smartphones on the market.
\end{enumerate}

\subsection{Toolchain}

The bytecode instrumentation process features the following steps:
1) Extract code from Android application apk files;
2) Modify the extracted code with bytecode manipulation tools;
3) Rebuild a new Android application containing the modified code.  

Those three steps can be broken down into five elementary steps, as shown in Figure \ref{fig:generateNewApk}: 
i) Extracting and converting the Dalvik bytecode into Java bytecode (step a-b), 
ii) Manipulating the bytecode (steps b-c),
iii) Translating this representation back to Dalvik bytecode  (step c-d), 
iv) Rebuilding a new apk file (step d-e) and 
v) Finally signing all files with a new private key (step e-f). 
Let us now discuss the tools that are used in each step.

\paragraph{i) Extracting the Dalvik Bytecode} The first step, as shown in 
Fig. \ref{fig:generateNewApk}.(a-b), is to extract the 
\texttt{classes\nolinebreak[0].dex} file 
from the apk file and convert it to Java bytecode classes which can be analyzed with standard unmodified Java bytecode analysis toolkits. 
For this step, we use the tool \texttt{dex2jar}\footnote{available at \url{http://code.google.com/p/dex2jar/}}. 

\paragraph{ii) Instrumenting the Bytecode} 
In this step, we experiment with two different tools which manipulate bytecode.
Recall that bytecode manipulation is the step from (b) to (c) at illustrated in Figure \ref{fig:generateNewApk}.
Using different tools gives us the opportunity to measure the difference in terms of execution time and memory consumption between them and decide which one is more appropriate to manipulate bytecode in a memory constrained system.
 
\emph{ii.a) Soot}. Classes are transformed to Jimple with the Soot analysis toolkit.
Soot \cite{lam2011soot} is an open-source analysis toolkit for Java programs. 
It operates either on Java source code or bytecode. 
It allows developers to analyze and transform programs. 
For instance, an intra-procedural flow analysis could determine if a variable can be \emph{null} at some point in the code.
Soot can also perform different call-graph analyzes, useful for specific bytecode instrumentation.
Most analyses and transformations in Soot use an internal representation called Jimple.
Jimple is a simple stack-less representation of Java bytecode.
We ported Soot on the Android system by converting its Java bytecode to Dalvik and creating a wrapper Android application.
To our knowledge there is no previous work which represent Android bytecode as an abstraction on which on could perform static analysis directly on the smartphone.

\emph{ii.b) ASM}. We experienced that Soot is sometimes slow and requires a lot of resources (especially memory). Thus, we also
run ASM for bytecode instrumentation.
ASM \cite{ASM} is a Java bytecode engineering library. One of its characteristics is that it is lightweight hence more suitable for running on systems constrained in term of memory or processing resource. 
It is primarily designed to manipulate and transform bytecode although it can also be used to perform some program analysis. It features a core API to perform simple transformations as well as a tree API to perform more complex bytecode transformations (which requires more CPU processing and memory space).

\paragraph{iii) Translating the Modified Bytecode back to Dalvik Bytecode}
Once the classes are analyzed and modified by the analysis toolkit, they are transformed back into Dalvik bytecode using \texttt{dx}\footnote{using \texttt{\small com.android.dx.command.Main} from the Android SDK} which generates the \texttt{classes.dex} file from Java class files. 
This step is illustrated in Fig. \ref{fig:generateNewApk} as the edge c-d.

\paragraph{iv) Rebuilding Application} As presented in Fig. \ref{fig:generateNewApk}.(d-e), after the fourth step, a new Andoid application is built. 
The newly generated \texttt{classes.dex}, the data and the Android manifest from the original application are all inserted in a new zip\footnote{using the \texttt{\small java.util.zip} library} file.

\paragraph{v) Signing the Modified Application} 
Android requires applications to be cryptographically signed.
Hence, all files of the generated zip file are signed using a newly created couple of public/private keys (not represented on the figure), 
The new public key is added to the zip (not represented on the figure). 
We used the \texttt{keytool} and \texttt{jarsigner} Java programs to sign applications (Fig. \ref{fig:generateNewApk}.(e-f)).

Signing applications with new keys may cause compatibility problems between applications.
For instance two or more applications signed with the same key can share the same process.
In order for this feature to continue working a one-to-one mapping between old keys and new ones needs to be maintained in order to sign two transformed applications (originally signed with the same keys) with the same new generated keys.
Maintaining this mapping and handling such compatibility between applications is out of scope of this paper.

 \paragraph{}
We have devised a bytecode manipulation process on Android using standard tools. 
The following presents the design and implementation of two concrete bytecode instrumentation prototypes.

\section{Use-case Design and Implementation}
\label{sec:useCases-implem}

Any use-case leveraging the toolchain presented in Section \ref{sec:methodology} analyzes or modifies the bytecode of an application.
Analyzing or modifying the bytecode is represented by step (b-c) in Figure \ref{fig:generateNewApk}.
We now present how we have implemented and evaluated the two use-cases of Section \ref{sec:useCases}.
Thy both modify the bytecode of applications. 
AdRemover modifies the bytecode to remove advertisement.
\ucperm{} modifies the bytecode to enable a fine-grained permission policy system for the user.

\subsection{Implementation of AdRemover}
\label{sec:implem-AdRemover}

We focus on two widely used Android advertisement modules: AdMob and AdSense.
Advertisement is not part of the Android system but is present in the application's bytecode.
Thus, applications do not share ad library code.
However, they each have a copy of the library code.
Disabling advertisement requires to instrument every application containing an ad library.

Advertisement requires I/O operations for fetching the ad data. 
An Android application developer using an ad library do not want her app to crash because of the ad library.
This is the reason why developers of ad libraries take special care of exceptions when designing the ad library.
They expect I/O operations to fail on a regular basis, depending on unpredictable contexts. 
For example, an exception can be thrown if the device has no network coverage anymore. 

Building on this observation, we make the assumption that I/O code has been placed by ad developers inside a Try/Catch block to recover for exceptions raised by I/O failures. 
Our tool leverages this assumption and inhibits every Try/Catch section of the ad packages of the application. 
For every Try/Catch block it encounters, our tool extracts the type of the handled I/O exception, creates such an exception object, and inserts an instruction that throws this exception at the very beginning of the try block. 

For this, we collected the Java package names used by these libraries and we configured AdRemover to operate only on classes that are part of those packages. 
We wrote two implementations of AdRemover: One using Soot and one using ASM.

\subsection{\ucperm{}: A Fine-grained Permission Policy Management}

In this context a fine-grained policy is a file in which the user specifies which permissions are granted to applications.
In the real world users are only familiar with permissions and applications, so it makes perfect sense to limit policies at the level of applications and not a lower level (such as Android component or Java methods).
However, for explanatory purposes the policies in this Section contain a mapping between Java methods and permissions.

For a user-centric policy to exist, we need to instrument the bytecode of every application one wishes to control.  
Recall from Section \ref{sec:android-in-brief} that Android applications communicate to the Android system through the Android API.
The instrumentation detects all API calls protected by one or more permissions and redirected every of those calls to a \emph{policy service}.
The policy service is a Android service component part of independent Android application.
Base on the user defined policy it authorizes or not the application to call the protected method.

When the instrumented application runs, the user-defined policy is enforced by the policy service.
Indeed, for every instrumented method, the running instrumented application calls the policy service and the policy is checked. 
If the policy allows the original API method call, the API call is performed. 
Otherwise, a fake implementation is executed and returns a fake default value.

Our prototype tool enforces a user-defined policy at the user level (also called application level).
It allows users who previously could not modify the system policy to enforce their own policy for a set of applications.
Modifying code to insert security check is known as Inline Reference Monitoring (IRM) and has been first introduced by Erlingsson et al. and Evans et al. \cite{erlingsson2003inlined, irm-00, naccio}.

\paragraph{Instrumenting the Application}

To control or limit an application's permission it's bytecode has to be instrumented.
This is illustrated in Figure \ref{fig:step1} where application $\appName{}$ is represented as a graph of method calls starting from node $s$. 
All method calls that require one or more permissions \cite{Felt2011a,bartel2012automatically} are wrapped with code which in order:
\begin{enumerate}\itemsep0em
\item asks the policy service if the application is authorized to call the method
\item according to the answer from the policy service either invokes the original method or the fake method.
\end{enumerate}

For instance, the \texttt{get\nolinebreak[0]Location(p1)} method invocation of node \emph{7} (which requires permission \texttt{GPS}) has been wrapped in the figure by a call to the \emph{policy} service.
If the policy approves this call, the original \texttt{get\nolinebreak[0]Location(p1)} is executed, otherwise a fake method is invoked, returning a fake default value.

In total, there are $N$ instrumentations where $N$ is the number of API calls under consideration present
in the application bytecode.

\paragraph{Defining the Policy}
The next step, as shown in Figure \ref{fig:step2}, is to define the policy regarding the instrumented applications.
The user defines a set of allowed permissions for each application.
Behind the scene, the policy generates a list of all Java methods which require the enabled permissions.
Those methods are set as authorized.
In Figure \ref{fig:step2}, only method \texttt{getLocation()} is allowed for application \instrumentedAppName{}. 

Note that this step could be performed first to instrument only method calls which are not authorized by the policy. 
However, instrumenting every API method calls which requires one or more permissions makes it possible to change the policy at runtime.

\begin{figure}[!]
\begin{center}
\resizebox{\columnwidth}{!}{

\newcommand{\UCFPSOFFSET}{5.4}
\newcommand{\POLICYOFFSET}{5.6}

\pgfdeclarelayer{background}
\pgfdeclarelayer{foreground}
\pgfsetlayers{background,main,foreground}
\tikzstyle{block} = [draw,fill=white!20,minimum width=4em]
\tikzstyle{steps} = [circle, draw, above, inner sep=1pt, midway, yshift=.2cm, fill=black!50, text=white]
\begin{tikzpicture} [node distance=.5cm, start chain=going below, remember picture]

		  \tikzstyle{vertex}=[circle,fill=black!25,minimum size=14pt,inner sep=0pt]
		  \tikzstyle{vInterface}=[block,fill=black!25,minimum size=14pt,inner sep=0pt]
		  \tikzstyle{vPermission}=[diamond,fill=blue!25,minimum size=14pt,inner sep=0pt]
		  \tikzstyle{bytecode}=[]
  		\tikzstyle{tuborg}=[decorate]
  		\tikzstyle{tubnode}=[midway, right=2pt]
			\tikzstyle{stub} = [rectangle, fill=yellow!5, text width=1.8cm, minimum height=.5cm, node distance=.6cm, draw];
		  \foreach \name/\x/\y in {s/1/0, 2/0/1, 3/1/1, 4/2/1, 5/1/2}
		    \node[vertex] (G-\name) at (\x,-\y/1.2) {$\name$};
		  \node[vertex] (G-e1) at (0,-3/1.2) {6};
		  \node[vertex, fill=red!40, draw] (G-e2) at (1,-3/1.2) {7};
		  \node[vertex] (G-e3) at (2,-3/1.2) {8};
		  \foreach \from/\to in {s/2, s/3, s/e3, 2/e1, 2/3, 3/5, 3/4, 5/2, 5/4, 5/e2}
		    \draw[->] (G-\from) -- (G-\to);
			\node[bytecode, draw, fill=red!40] (GC-bt7) at (1, -4.1) {\footnotesize 
				 \begin{minipage}[t][1.6cm]{4.3cm}
					\tikz[remember picture] \coordinate (C-toStub); \\
						\phantom{}if (policyAccepts(getLocation);\tikz[remember picture] \coordinate (C-policyCheck);)\\
							\phantom{ee}r = obj.getLocation(p1);\tikz[remember picture] \coordinate (C-execA);\\
						\phantom{}else\\
							\phantom{ee}\tikz[remember picture] \coordinate (C-execBleft); r = stub.getLocation(p1);\tikz[remember picture] \coordinate (C-execB);\\ 
					\end{minipage} 
						};

			\node[rectangle, text width = 4cm] (G-Ucode) at (\UCFPSOFFSET+1.5, -2/1.15) {
				 \begin{minipage}[t][4cm]{4cm}
					\tikz[remember picture] \coordinate (C-toStub); 
policyAccepts(Method m) \\
						\phantom{ee}if (policyHas(m)\tikz[remember picture] \coordinate (C-policyCheck);)\\
							\phantom{eee}return True;\tikz[remember picture] \coordinate (C-execA);\\
						\phantom{ee}else\\
							\phantom{eee}\tikz[remember picture] \coordinate (C-execBleft);return False;\tikz[remember picture] \coordinate (C-execB);\\ 
					\end{minipage} 
				};
				\node[stub, below of=G-Ucode] (stub1) {code stub 1};
				\node[stub, below of=stub1] (stub2) {code stub 2};
				\node[stub, below of=stub2] (stubdots) {$\hdots$};
				\node[stub, below of=stubdots] (stubn) {code stub N};

		  \node[draw, text width=2.3cm, minimum height=2.8cm, label=above:Policy file] (GU-policy) at (\UCFPSOFFSET+\POLICYOFFSET,-2/1.2) {
				\tikz[remember picture] \coordinate (C-inPolicy);\instrumentedAppName{} \{\\
					\hspace{.2cm}getLocation();\\ 
				\}\\
			};

\begin{pgfonlayer}{background}
		  	\node[] (G-00) at (0,0) {};
		  	\node[] (G-10) at (2,0) {};
		  	\node[] (G-01) at (0,-3/1.2) {};
		  	\node[] (G-11) at (2,-3/1.2) {};
		  	\node[] (G-U00) at (0+\UCFPSOFFSET,0) {};
		  	\node[] (G-U10) at (3+\UCFPSOFFSET,0) {};
		  	\node[] (G-U01) at (0+\UCFPSOFFSET,-5/1.2) {};
		  	\node[] (G-U11) at (3+\UCFPSOFFSET,-5/1.2) {};
        \path (G-00.west |- G-10.north)+(-0.3,0.3) node (a) {};
        \path (G-01.north -| G-11.east)+(0.3,-0.5) node (b) {} coordinate (cright);
        \path[fill=black!5,rounded corners, draw=black!50, dashed]
            (a) rectangle (b);
        \path (G-U00.west |- G-U10.north)+(-0.4,0.3) node (a) {};
        \path (G-U01.north -| G-U11.east)+(0.4,-0.5) node (b) {};
        \path[fill=black!5,rounded corners, draw=black!50, dashed]
            (a) rectangle (b);
        \path (G-U00.west |- G-U10.north)+(-0.3,0.2) node (c) {};
        \path (G-U01.north -| G-U11.east)+(0.3,2.2) node (d) {};
        \path[fill=yellow!5,rounded corners, draw]
            (c) rectangle (d);
\end{pgfonlayer}

			\draw[-latex, dashed, color=red!40] (G-e2) -- (GC-bt7);

			\draw[tuborg, decoration={brace}] let
			    \p1=(G-00.west), \p2=(G-10.east) in
					    ($(\x1-.3cm,.6)$) -- ($(\x2+.3cm,.6)$) node[tubnode, above]  {\instrumentedAppName{}$\hspace{.3cm}$};
			\draw[tuborg, decoration={brace}] let
			    \p1=(G-U00.west), \p2=(G-U10.east) in
					    ($(\x1-.3cm,.6)$) -- ($(\x2+.3cm,.6)$) node[tubnode, above]  {Policy service$\hspace{.3cm}$};

			\draw[-latex] (GC-bt7.east) -- +(1,0) node[steps] {A} |- (C-toStub);
			\draw[-latex] (C-policyCheck) -- node[steps] {B} (C-inPolicy); 
			\path[] (C-execA) -- node[steps, yshift=-.5cm, xshift=.3cm] {C} (C-execA); 
		     
\end{tikzpicture}
}
\caption{\label{fig:step3}Step 3: the Policy Monitor Enforces the Fine-Grained Permissions by Returning Default Values for Unauthorized API Calls}
\end{center}
\end{figure}
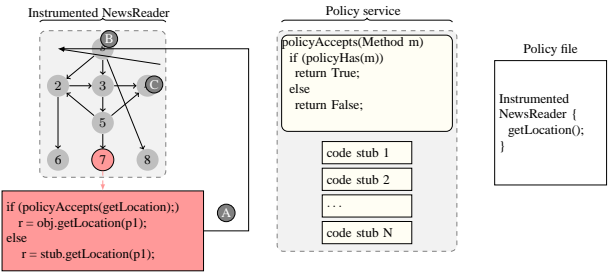

\paragraph{Policy Service}
Finally, when the instrumented application runs, the policy is enforced by the Policy service as shown in Figure \ref{fig:step3}. 
For every instrumented method (here the original/instrumented method is \texttt{get\nolinebreak[0]Location} and its associated permission \texttt{GPS}) 
the running application calls \texttt{policy\nolinebreak[0]Accepts()} (step \emph{A}) and the policy is checked by calling\linebreak\texttt{policyHas()} (step \emph{B}). 
Method \texttt{policy\nolinebreak[0]Accepts()} returns \texttt{true} if the policy allows the original method or false if it does not.
If the original method is allowed in the policy, the original method is called (this is the case in Figure \ref{fig:step3}, since step \emph{C} returns \emph{true}). 
Otherwise, the stub method corresponding to the original method is executed. 
Here, the stub handling method \texttt{getLocation} is not executed.
We implemented the policy service as an Android service and the instrumentation code as a plugin for the static analysis tool Soot.


\begin{figure}[!]
\begin{center}
\resizebox{\columnwidth}{!}{

\newcommand{\UCFPSOFFSET}{5.4}

\pgfdeclarelayer{background}
\pgfdeclarelayer{foreground}
\pgfsetlayers{background,main,foreground}
\tikzstyle{block} = [draw,fill=white!20,minimum width=4em]
\begin{tikzpicture}
  [node distance=.5cm,
  start chain=going below,]

		  \tikzstyle{vertex}=[circle,fill=black!25,minimum size=14pt,inner sep=0pt]
		  \tikzstyle{vInterface}=[block,fill=black!25,minimum size=14pt,inner sep=0pt]
		  \tikzstyle{vPermission}=[diamond,fill=blue!25,minimum size=14pt,inner sep=0pt]
		  \tikzstyle{bytecode}=[]
  		\tikzstyle{tuborg}=[decorate]
  		\tikzstyle{tubnode}=[midway, right=2pt]
		  \foreach \name/\x/\y in {s/1/0, 2/0/1, 3/1/1, 4/2/1, 5/1/2}
		    \node[vertex] (G-\name) at (\x,-\y/1.2) {$\name$};
		  \node[vertex] (G-e1) at (0,-3/1.2) {6};
		  \node[vertex, fill=green!40, draw] (G-e2) at (1,-3/1.2) {7};
		  \node[vertex] (G-e3) at (2,-3/1.2) {8};
		  \foreach \from/\to in {s/2, s/3, s/e3, 2/e1, 2/3, 3/5, 3/4, 5/2, 5/4, 5/e2}
		    \draw[->] (G-\from) -- (G-\to);
			\node[bytecode, text width=3.5cm, draw, fill=green!40] (G-bt7) at (1, -4.1) {\footnotesize r = obj.getLocation(p1);};

		  \foreach \name/\x/\y in {s/1/0, 2/0/1, 3/1/1, 4/2/1, 5/1/2}
		    \node[vertex] (GU-\name) at (\x+\UCFPSOFFSET,-\y/1.2) {$\name$};
		  \node[vertex] (GU-e1) at (0+\UCFPSOFFSET,-3/1.2) {6};
		  \node[vertex, fill=red!40, draw] (GU-e2) at (1+\UCFPSOFFSET,-3/1.2) {7};
		  \node[vertex] (GU-e3) at (2+\UCFPSOFFSET,-3/1.2) {8};
		  \foreach \from/\to in {s/2, s/3, s/e3, 2/e1, 2/3, 3/5, 3/4, 5/2, 5/4, 5/e2}
		    \draw[->] (GU-\from) -- (GU-\to);
			\node[bytecode, draw, fill=red!40] (GU-bt7) at (.5+\UCFPSOFFSET, -4.1) {\footnotesize 
				 \begin{minipage}[t][1.6cm]{4.3cm}
					\tikz[remember picture] \coordinate (C-toStub);
						\phantom{}if (policyAccepts(getLocation)\tikz[remember picture] \coordinate (C-policyCheck);)\\
							\phantom{ee}r = obj.getLocation(p1);\tikz[remember picture] \coordinate (C-execA);\\
						\phantom{}else\\
							\phantom{ee}\tikz[remember picture] \coordinate (C-execBleft); r = stub.getLocation(p1);\tikz[remember picture] \coordinate (C-execB);\\ 
					\end{minipage} 
				};

\begin{pgfonlayer}{background}
		  	\node[] (G-00) at (0,0) {};
		  	\node[] (G-10) at (2,0) {};
		  	\node[] (G-01) at (0,-3/1.2) {};
		  	\node[] (G-11) at (2,-3/1.2) {};
		  	\node[] (G-U00) at (0+\UCFPSOFFSET,0) {};
		  	\node[] (G-U10) at (2+\UCFPSOFFSET,0) {};
		  	\node[] (G-U01) at (0+\UCFPSOFFSET,-3/1.2) {};
		  	\node[] (G-U11) at (2+\UCFPSOFFSET,-3/1.2) {};
        \path (G-00.west |- G-10.north)+(-0.3,0.3) node (a) {};
        \path (G-01.north -| G-11.east)+(0.3,-0.5) node (b) {} coordinate (cright);
        \path[fill=black!5,rounded corners, draw=black!50, dashed]
            (a) rectangle (b);
        \path (G-U00.west |- G-U10.north)+(-0.3,0.3) node (a) {} coordinate (cleft);
        \path (G-U01.north -| G-U11.east)+(0.3,-0.5) node (b) {};
        \path[fill=black!5,rounded corners, draw=black!50, dashed]
            (a) rectangle (b);
\end{pgfonlayer}
			\draw[-latex] let
				\p1=(cright), \p2=(cleft) in
					(\x1+.1cm, -1.1) -- (\x2-.1cm, -1.1) node[above, midway] {step1};

			\draw[-latex, dashed, color=green!40] (G-e2) -- (G-bt7);
			\draw[-latex, dashed, color=red!40] (GU-e2) -- (GU-bt7);

			\draw[tuborg, decoration={brace}] let
			    \p1=(G-00.west), \p2=(G-10.east) in
					    ($(\x1-.3cm,.6)$) -- ($(\x2+.3cm,.6)$) node[tubnode, above]  {\appName{}$\hspace{.3cm}$};
			\draw[tuborg, decoration={brace}] let
			    \p1=(G-U00.west), \p2=(G-U10.east) in
					    ($(\x1-.3cm,.6)$) -- ($(\x2+.3cm,.6)$) node[tubnode, above]  {\instrumentedAppName{}$\hspace{.3cm}$};

\end{tikzpicture}
}
\caption{\label{fig:step1}Step 1: Wrapping and Redirection of Android API Calls For Fine-Grained Permission Management}
\end{center}
\end{figure}
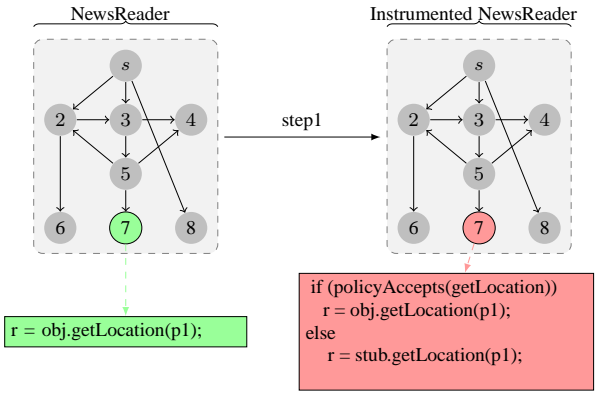

\begin{figure}[!]
\begin{center}

\pgfdeclarelayer{background}
\pgfdeclarelayer{foreground}
\pgfsetlayers{background,main,foreground}
\tikzstyle{block} = [draw,fill=white!20,minimum width=4em]
\begin{tikzpicture}
  [node distance=.5cm,
  start chain=going below,]

		  \node[draw, text width=4cm, minimum height=2.8cm, label=above:Policy file] (GU-policy) at (0,0) {
				\instrumentedAppName{} \{\\
					\hspace{.2cm}getLocation();\\ 
				\}\\
			};

\end{tikzpicture}
\caption{\label{fig:step2}Step 2: The Policy File Defines that InstrumentedNewsReader is Allowed to Use API Method getLocation}
\end{center}
\end{figure}
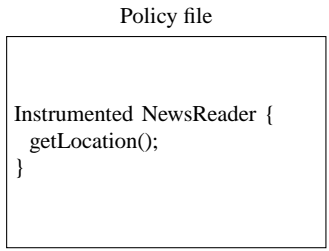

\subsection{Evaluation}

We now check whether our use-case implementations work. 
For both of them, we run the instrumentation against a real-world application and runs the resulting modified application.

\paragraph{\ucad}
We test that our tool is functional by selecting a random application on the Android Market.
We make sure that the test application uses one of the two advertisement modules currently handled by \ucad. 

First we run the unmodified test application on an Android devices, and make sure that it is a working application, and that it actually displays advertisements. 

We then send this application to our toolchain (with the Soot implementation) running on a PC.
The modified application is still functional, and no more advertisements are displayed. 
We monitor the network connection during the test and found out that it the application does not send any ad request anymore. 

Finally, we process the unmodified application again, this time running the bytecode manipulation directly on the smartphone. 
Running the modified application yielded the same results as with the application modified on a standard PC. 

\paragraph{\ucperm}
For evaluating the fine-grain policy, we select another random application and instrument it to wrap every permission sensitive API call related to the GPS. 
The application is instrumented and then repackaged into a new signed application. 
We run the instrumented application on an Android device, and test it with different policies. 
The user-defined policy is enforced as expected.

\paragraph{}
To sum up, the two bytecode transformations result in applications that correctly runs. 
Those first results are important as the two use cases illustrate what can be achieved using the bytecode instrumentation toolchain.
What also matters for us is to know whether the toolchain under consideration can be run in vivo on a large dataset of Android applications given the memory and CPU limitations of current smartphones.
The next section answers to this questions by measuring execution time and memory consumption of in vivo instrumentation.


\section{Performance of In-Vivo Instrumentation}
\label{sec:eval}

In this section we present the results of applying the instrumentation process presented in Section \ref{sec:methodology} and summarized in Fig. \ref{fig:generateNewApk}. 
The goal is to know:
1) whether it is possible to manipulate bytecode on smartphones given the restricted resources of the hardware.
2) whether it takes a reasonable amount of time.

\subsection{Measures}\label{subsec:measures}

We measure  the execution time of the five steps of the instrumentation process on a set of 130 Android applications.
This set is described in Section \ref{sec:dataset}. 
We run the instrumentation process on three different Android smartphones whose configurations are presented in Section \ref{sec:material}. 

The feasibility of the whole process is measured by the time to pass every step of the toolchain (1: $dex2jar$, 2: $Soot/ASM$, 3: $dx$, 4: $customZip$, 5: $signature$). 
The time to run each step and the number of applications that successfully go through each step are measured as well.

For the second step of the process (Step: Instrumenting the bytecode), we evaluate both $ASM$ and $Soot$.  
For $ASM$, we measure the time required to instrument Java bytecode on the \ucad{} case study. 
The \ucad{} transformation leverages the ASM tree API to perform the try/catch block manipulation described in \ref{sec:implem-AdRemover}.
$Soot$ is evaluated by measuring the time required to generate Java classes for both \ucad{} and \ucperm{} case studies (\ucad{} is implemented with ASM and Soot).

\subsection{\label{sec:material}Experimental Material}

We conduct the experiment on three Android-based smartphone devices. 
Their configuration is detailed in Table \ref{table:configurations}.
The main differences are the processor clock speed (0.8, 1.2 and 1.4 GHz), the total amount of main memory (512, 768 and 1024 MiB), the Android version (2.2, 2.3.4 and 4.0.3) and the  maximum heap size of the Dalvik virtual machine (24, 32 and 48). 
Since the heap size controls the maximum memory that can be allocated by a single process it also controls the maximum number of objects that can be allocated simultaneously.

The number of cores also differs. 
However, we do not take advantage of multiple cores during the experiments.
This hardware complies with requirement \#3 mentioned in \ref{reqs}.

\begin{table}
\begin{center}
\resizebox{\columnwidth}{!}{
  \begin{tabular}{|l|l|r|l|r|}
  \hline
  Name    &   Processor   & Memory  & Android                           & Heap Size                                   \\ \hline 
  smartphone1 & ARM 800MHz, 1 core  & 512MiB  & 2.2   & 24MiB     \\ \hline 
  smarthpone2 & ARM 1.2GHz, 2 cores & 768MiB  & 2.3.4 & 32MiB     \\ \hline 
  tablet1     & ARM 1.4GHz, 4 cores & 1GiB    & 4.0.3 & 48MiB     \\ \hline
  \end{tabular}
}
  \caption{\label{table:configurations}The Hardware used in our Experiment}
\end{center}
\end{table}

\subsection{\label{sec:dataset}Dataset}

We apply the whole experimental protocol on a set of 130 Android applications randomly selected among the top 500 applications from the Android market\footnote{\url{http://play.google.com}}.  
They span various domains such as finance, games, communications, multimedia, system or news.
This dataset is not artificial as it only consists of real world applications.

\begin{figure}
\begin{center}
\resizebox{\columnwidth}{!}{
\input{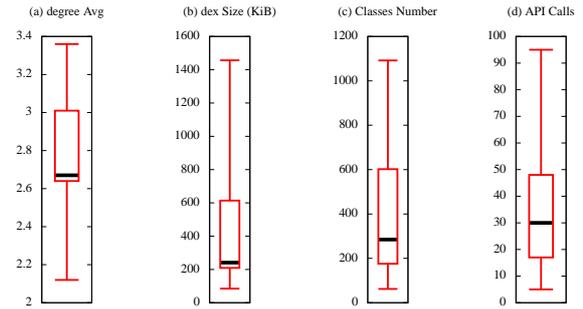}
}
  \caption{\label{fig:statsApk}Descriptive Statistics of the 130 Applications of our Dataset}
\end{center}
\end{figure}

To give a better overview on these applications, Figure \ref{fig:statsApk} shows the key application metrics as boxplots. 
They indicate that most (75\%) of Android applications have less than 614 KiB of Dalvik bytecode, less than 602 classes, an average method degree smaller than 3. 
Haft of the applications have more than 30 calls to a method of the Android API which require a permission.

\begin{figure}
\begin{center}
\resizebox{\columnwidth}{!}{
\input{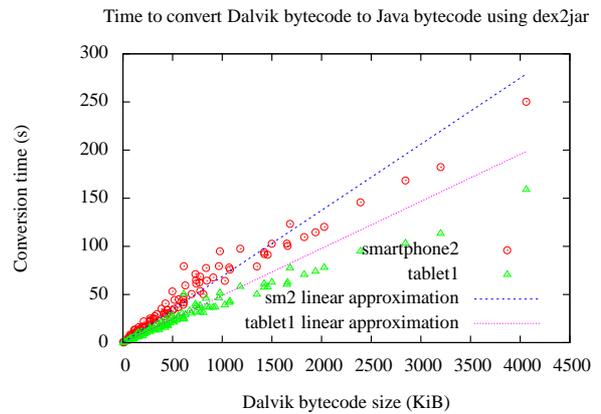}}
\caption{\label{fig:dex2jarTime}Performance of Dalvik to In-Vivo Java Bytecode Conversion.}
\end{center}
\end{figure}

\subsection{\label{subsec:dex2jar}Dalvik to Java Bytecode Conversion}

The conversion time from the Dalvik executable code to Java bytecode using \emph{dex2jar} is shown in Fig. \ref{fig:dex2jarTime}.

\observation The time to convert dex files to jar does not exceed 60 seconds on \spt{} and \tabone{} for 75\% of the applications. 
The conversion time does not exceed 250 seconds on our dataset of Android applications.

\observation The application with the biggest Dalvik bytecode file (4000 KiB) is successfully converted both on \spt{} and on \tabone{}.

\observation We notice that the conversion time is linear with the size of the dex file 
(of the form $a \cdot X + b$) for a Dalvik bytecode size less than 4000 KiB. 
Using linear regression, we find that for \spt{} $a$ equals 0.069 and $b$ equals 0.3. 
For \tabone{} we have, 0.049 and -0.4. 
The linear relation between the conversion time and the size of the Dalvik bytecode enables us to theoretically predict
the necessary amount of time to convert any size of Dalvik bytecode (if we extrapolate for size bigger than 4000 KiB). 
For instance, the time to process the Android application with 10 MiB of Dalvik bytecode would be 700 seconds for \spt{} and 500 seconds for \tabone{}. 

\conclusion Converting Dalvik bytecode to Java bytecode in-vivo is feasible  within minutes.

\emph{Limitations:} \spo{} is unable to process any dex file.
Also, when using \spt{} and \tabone{}, 26 and 11 dex files, respectively, cause the conversion Android application dex2jar to crash. 
This crash is either an \texttt{Out\-Of\-Memory} or a \texttt{Stack\-Overflow} exception. 

Result of \spo{} is explained by the hard-coded maximum heap size of Android (32 MiB or 48 MiB).
For the two other devices, crashes are to be attributed to the default 8 KiB stack size.
In total, 104 (80\%) Android applications were successfully converted to a jar file on \spt{} and  119 (91\%) on \tabone{}.

However, since Android devices become more and more powerful the default heap size of the Android system grows. 
Indeed, in Android 2.2 the heap size is 24 MiB, in Android 2.3.4 32 MiB and in Android 3.0 48 MiB.
This continued growth would allow our tool chain to convert Android applications which have bigger Dalvik bytecode size.

Also, some applications may be obfuscated to prevent Dex2jar to convert Dalvik bytecode to Java classes.
We did not encounter any obfuscation during the experiment. 
Our toolchain relies on independent components. 
Thus, if Dex2jar cannot handle some obfuscation techniques it could easily be replaced by an equivalent component which handles them.


\subsection{Performance of Bytecode Manipulation}

This section presents our performance measures of in-vivo bytecode manipulation using two different instrumentation libraries: ASM and Soot.

\subsubsection{Manipulation With ASM}
\begin{figure}
\begin{center}
\resizebox{\columnwidth}{!}{
	\input{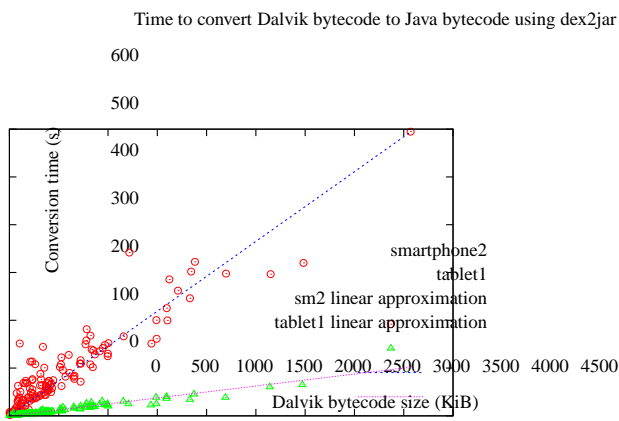}
}

\caption{\label{fig:ASMTime}Transformation Time of In-Vivo Java Bytecode Manipulation with ASM}
\end{center}
\end{figure}

Transformation time of Java bytecode using ASM is represented in Figure \ref{fig:ASMTime}.
In this experiment the \ucad{} transformation described in \ref{sec:adremove} is implemented using ASM.

\observation All 104 applications successfully transformed with dex2jar on \spt{} are successfully processed by ASM in-vivo. 
It processes every jar (up to 4MiB in size) in less than 600 seconds.

\observation We notice that the transformation time is linear with the size of the jar files (of the form $a \cdot X + b$) for a Dalvik bytecode size less than 4000 KiB. 
Using linear regression, we find that for \spt{} $a$ equals 0.146.
For \tabone{} we have, 0.025. 

\conclusion Manipulating bytecode on smartphones using ASM is feasible.
Given our transformation and our dataset, ASM does not have specific memory or CPU requirements that are incompatible with smartphone resources.

\subsubsection{Manipulation With Soot}

We now consider the Soot implementation of the AdRemover transformation.
Out of the 130 Android applications, only 3/130 are correctly processed on \spt{} and 19/130 are correctly processed on \tabone{}. 

\observation Only the smallest applications (in terms of Dalvik bytecode) can be converted.
For instance, it takes less than 30 seconds to convert any jar which size is less or equal to 20 KiB on \spt{}.
However, larger, yet small applications (in the 25\% quartile), take up to 18 minutes for being instrumented with Soot. 

\begin{figure}
\begin{center}
\resizebox{\columnwidth}{!}{
\input{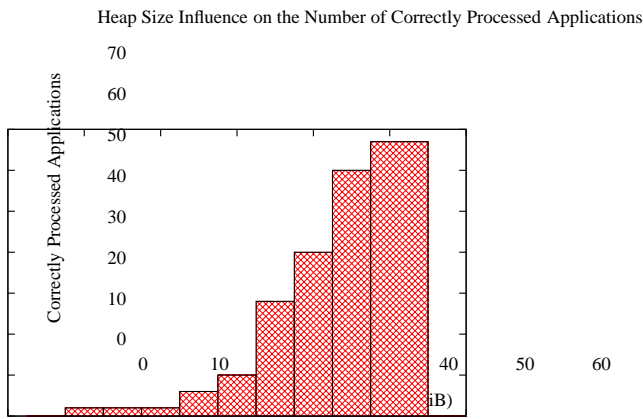}
}
\caption{\label{fig:heapSize}Influence of the Heap Size on Jimple Transformation}
\end{center}
\end{figure}

\conclusion 
Using Soot in-vivo is feasible only for the smallest applications.
We assume that the heap size is the main blocking factor of using Soot in-vivo.
To check this assumption, we conducted an experiment on a desktop computer consisting of analyzing our dataset of Android applications with 
different maximal heap sizes (from 5 Mib to 50 Mib by steps of 5 Mib). Results are presented Fig. \ref{fig:heapSize}. 
Soot was able to process 67 applications with a heap size of 50 Mib. 
Those results clearly indicate that maximum half of the Android applications could be processed with a heap size of 50 MiB. 
Under the assumption that the heap usage (hence the maximum required size) is similar on the Java and Dalvik virtual machines, it means that the memory is actually the main blocking factor of using Soot on Android.

\subsection{Java Bytecode to Dalvik Conversion}

\begin{figure}
\begin{center}
\resizebox{\columnwidth}{!}{
\input{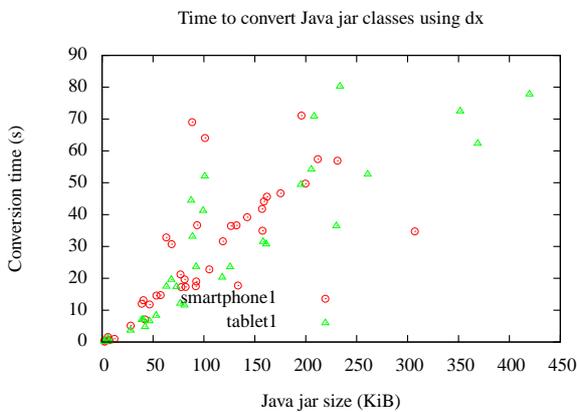}
}

\caption{\label{fig:dxTime}Conversion Time of In-Vivo  Java Bytecode to Dalvik Translation.}
\end{center}
\end{figure}

Once an application has been instrumented at the Java bytecode level, it has to be transformed back into Dalvik bytecode.
Conversion time from Java classes to the dex file using the \emph{dx} tool is shown in Fig. \ref{fig:dxTime}.

\observation Java bytecode of 33/130 on \spt{} and and 39/130 applications on \tabone{}, respectively,  have been successfully converted to Dalvik bytecode.  

\observation Conversion time for jar files ranging from 20 to 400 KiB does not exceed 80 seconds. 

\conclusion
The Dx tool is a bottleneck of the tool chain. It can only correctly process 25 to 30\% of the applications.
The reason is that it puts every Java class in memory and suffers from the memory limitation of in-vivo processing, similarly to Soot.
This tool is used off the shelf and could be optimized to run on devices where resources are limited, by processing class after class to limit memory consumption.
.

\subsection{Creating a New \texttt{apk} File}

\begin{figure}
\begin{center}
\resizebox{\columnwidth}{!}{
\input{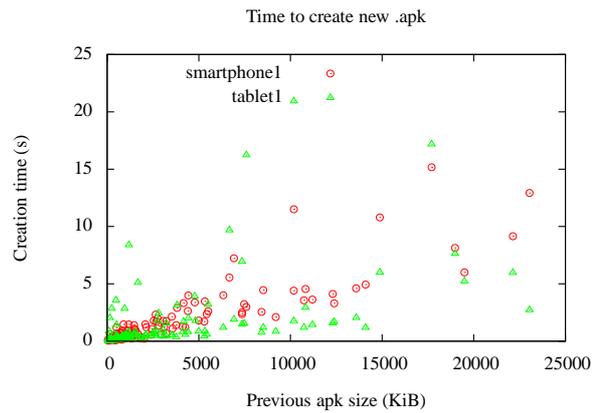}
}
\caption{\label{fig:apkTime}Creation Time of a New \texttt{apk} File In-Vivo }
\end{center}
\end{figure}

The time taken to create an \texttt{apk} file from the instrumented Dalvik bytecode is shown in Fig. \ref{fig:apkTime}.
Note that for this step, the input set is not the output of the previous step. 
We only have 39/130 applications that have been correctly processed in the previous steps.
At every step, some applications failed.
For the remaining 91/130 applications where the final instrumented Dalvik bytecode could not be computed, we take as input the original Dalvik dex file of the application. 
In this way, the problems of the previous step do not interfere with the results of this fourth step.   

\observation 121/130 inputs were successfully processed.
There is no clear relation between the size of the previous \texttt{apk} file and the creation time of the new \texttt{apk}. 
Only 9/130 applications generate an exception because their size is too big and can thus not be processed by the zip utility.

\observation For 95\% of the applications it takes less than five seconds regardless of the device and of the size of the original \texttt{apk} file. 

\conclusion 
It is feasible to create \texttt{apk} files on smartphones. The time to create a new \texttt{apk} file is negligible compared to the time to convert bytecode or to manipulate bytecode with Soot.

There is no linear relation with the Dalvik size as it is the case in Fig. \ref{fig:dex2jarTime} and \ref{fig:dxTime}.
This is probably due to the fact that when generating apk files, others factors than the bytecode size come into play, such as handling the media files (images, sound, etc.), which sometimes dominate the Dalvik bytecode size.

\subsection{Signing the Generated \texttt{apk} File}

\begin{figure}
\begin{center}
\resizebox{\columnwidth}{!}{
\input{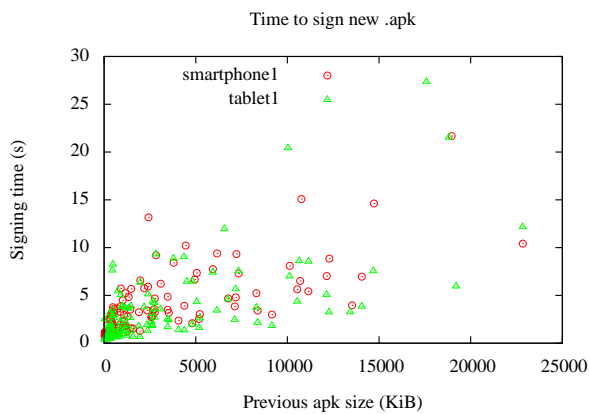}
}
\caption{\label{fig:apkSignedTime}Performance of In-Vivo Signature of the Instrumented Apk File}
\end{center}
\end{figure}

Signing time of applications is represented in Figure \ref{fig:apkSignedTime}. 

\observation 120/130 Android applications were successfully signed on \tabone{}. 
There is no clear relation between the size of the \texttt{apk} file and the signature time of the \texttt{apk} file. 
14/130 applications generate an exception because their size is too big and can thus not be processed (14 on \spt{} and 10 on \tabone{}).

\observation For 95\% of the applications a maximum of 12 seconds is required to sign the application regardless of the device and the size of the \texttt{apk} file.

\conclusion 
It is feasible to sign \texttt{apk} files on smartphones. 
Similarly to the \texttt{apk} file creation step, the computation time is negligible. 
The difference observed between \spo{} and \spt{} reflects the difference in their CPU clock frequencies.

\subsection{Conclusion}
We now recapitulate the results of our experiments of in-vivo modification of Android applications.

\subsubsection{Feasibility}

Table \ref{table:recapResults} summarizes all the experiments for \spt{} and highlights the feasibility of the whole approach.

Total execution times for all steps of the toolchain are computed for the Soot and ASM version.
For an ASM-based instrumentation it takes a median time of 120 seconds, that is 2 minutes, to process an application. 
We think that users would agree with waiting 2 minutes before starting using an application, if they are provided more guarantees with this instrumentation process enabling better privacy.
During those 2 minutes the phone is still usable since only once core is used (most smartphones feature multi-core CPUs) and only the maximum amount of heap memory allowed by the virtual machine can be used (and not all memory).

Those experiments show that it is feasible to manipulate bytecode directly on Android devices.
The most expensive steps of the process are the conversion of Dalvik to Java bytecode and vice versa, and the Soot bytecode manipulation step.

\begin{table*}
\begin{center}
	\begin{tabular}{|l|p{1.5cm}|p{1.5cm}|p{1.5cm}|p{1.5cm}|l|l|}
	\hline
	Step Name				&	Min. Time (s)	& Avg. Time (s) & Median Time (s)	&	Max. Time (s) 	&   App.		& Feasibility		\\ \hline \hline
	Conversion .dex to .jar (a-b)		&		0.22	&	43.76	&	28.9		&	250.2		& 104/130 (80\%)	&	$\star$$\star$$\star$  \\ \hline 
	Analyzing .jar with Soot(b-c) 		& 		25.8	& 	76  	&	26		& 	187.7		& 3/130 (2.3\%)		&			 		\\ \hline 
	Analyzing .jar with ASM(b-c) 		& 		1.55	& 	90.45	&	65.1		& 	594.67		& 129/130 (99.2\%)	&	$\star$$\star$$\star$$\star$		 		\\ \hline 
	Conversion class to dex	(c-d) 		& 		0.09	& 	28.07	&	22.8		& 	71  		& 39/130 (30\%)		&	$\star$$\star$	 		\\ \hline 
	Creating new .apk	(d-e)		& 		0.06	& 	1.89	&	0.87		& 	15.1		& 119/130 (91.5\%)	&	$\star$$\star$$\star$$\star$	 		\\ \hline 
	Signing new .apk	(e-f)		& 		0.71	& 	3.85	&	3.0 		& 	21.67		& 116/130 (89.2\%)	&	$\star$$\star$$\star$$\star$	 		\\ \hline \hline
	All Steps with Soot (a-b-c-d-e-f)	&		26.88	&	153.57	&	81.57		&	545.67		& 3/130 (2.3\%)	&	$\star$					\\	\hline
	All Steps with ASM (a-b-c-d-e-f)	&		2.63	&	168.02	&	120.67		&	952.64		& 39/130 (30\%)	&	$\star$$\star$$\star$					\\	\hline
	\end{tabular}
\caption{Summary Metrics of Our In-Vivo Instrumentation Process for Smartphone2. There are problematic steps but the overall process is feasible.}
\label{table:recapResults}
\end{center}
\end{table*}

\begin{table*}
\begin{center}
	\begin{tabular}{|l|p{1.5cm}|p{1.5cm}|p{1.5cm}|p{1.5cm}|l|l|}
	\hline
	Step Name				& Min. Time (s)	& Avg. Time (s) & Median Time (s)& Max. Time (s) &   App.						& Feasibility		\\ \hline \hline
	Conversion .dex to .jar (a-b)		&	0.19	&	25.6	&	17.85	& 158.9		&	119/130 (91.5\%)		&	$\star$$\star$$\star$  \\ \hline 
	Analyzing .jar with Soot(b-c) 		& 	24.2	& 	76	&	352	& 1054		& 19/130 (14.6\%)		&		$\star$	 		\\ \hline 
	Analyzing .jar with ASM(b-c) 		& 	1.55	& 	11.3	&	7.06	& 65.5  	& 119/130 (91.5\%)		&	$\star$$\star$$\star$$\star$		 		\\ \hline 
	Conversion class to dex	(c-d) 		& 	0.09	& 	29.5	&	20.2	& 80.2 		& 33/130 (25.3\%)		&	$\star$$\star$	 		\\ \hline 
	Creating new .apk	(d-e)		& 	0.03	& 	1.6	&	0.5 	& 20.9		& 121/130 (93.1\%)		&	$\star$$\star$$\star$$\star$	 		\\ \hline 
	Signing new .apk	(e-f)		& 	0.4 	& 	3.4 	&	1.91 	& 27.3		& 120/130 (92.3\%)		&	$\star$$\star$$\star$$\star$	 		\\ \hline \hline
	All Steps with Soot (a-b-c-d-e-f)	&	24.91	&	136.1	&	392.46	& 1341.3	& 19/130 (14.6\%)		&	$\star$					\\	\hline
	All Steps with ASM (a-b-c-d-e-f)	&	2.26	&	71.4	&	47.52	& 352.8		& 33/130 (25.3\%)		&	$\star$$\star$$\star$					\\	\hline
	\end{tabular}
\caption{Summary Metrics of Our In-Vivo Instrumentation Process for  (for Tablet1)}
\label{table:recapResults}
\end{center}
\end{table*}

\subsubsection{How to Improve Performance of In-Vivo Instrumentation?}
According to our analysis, the main blocking factor is the memory.
The maximum heap size required to analyze and transform applications is an issue for many transformation steps. 
We think that this issue can easily be solved by 1) the next generation of more powerful hardware and 2) the upcoming versions of the Android OS and virtual machines which will likely have a significantly higher maximum heap size (e.g. Android 4 heap size is set to 48 MiB). 

Dalvik to Java conversion and Java to Dalvik conversion are two very time expensive steps. 
They use unmodified versions of Dex2jar and dx.
There are two ways to overcome those resource-hungry tools.

First, those tools were never optimized to run on platforms with limited resources.
We believe that there are many optimization opportunities in terms of CPU and memory consumption.

Second, one could replace those tools by better alternatives.
For instance, an ASM-like library for manipulating Dalvik bytecode would allow to skip Dalvik-to-Java and Java-to-Dalvik conversion. 
Such tools are emerging such as ASMdex\footnote{See \url{http://asm.ow2.org/asmdex-index.html}}.
Another solution would consist of performing bi-directional transformations directly from Dalvik bytecode to Jimple bytecode which are both register based.
We are indeed working on a Dalvik to Jimple translation prototype called Dexpler \cite{bartel2012dexpler}.

To sum up, our results show that we can reasonably imagine to manipulate the bytecode on 100\% of our dataset applications within at most 5 minutes. 

\subsubsection{Threats to Validity}

Let us now discuss the threats to validity of our experimental results. 

\begin{description}\itemsep0em 

\item{Implementation Bug:}
Our results hold as far as there is no serious bug in the implementation of any of the five programs involved in the five steps, as well as in the glue and measurement code we wrote.

\item{Dataset Generalizability:} 
Our dataset may not be representative of the Android applications used in the real-world.

\item{Linear Extrapolation:}
The linear relations we establish for the Dalvik to Java and the Java to Dalvik conversions holds for bytecode size less or equal to 300 KiB. 
It may not hold for bytecode whose size is bigger. 
In the presence of non-linear singularities, it may not be possible to analyze large applications.

\item{Bytecode Manipulation Time:}
Our results on the bytecode manipulation time were obtained with relatively simple transformations. 
It may be the case that complex transformations are not of the same order of magnitude and consume much more memory.
However, for the use cases presented in Section \ref{sec:useCases}, the instrumentation only consists in monitoring and proxying Java methods. 

\end{description}


\section{Related Work}
\label{sec:related-work}


\paragraph{\bf Monitoring Applications}

Monitoring smart\-phone applications at runtime is an idea which recently emerged, due to the explosion of ``mobile'' malware and the increasing sophistication of mobile OS.

Bose et al. \cite{Bose2008} aimed at detecting malware based on their behavior at runtime.
For this, they added hooks in the Symbian OS emulator to track OS and API calls.
In other words, malware detection is only achieved in the emulator, \emph{in vitro}. 
On the contrary, we aim malware detection in live user environments, \emph{in vivo} and showed in this paper that is it is feasible in the mid-term.

Enck et al. \cite{Enck2010} presented a runtime monitoring framework called TaintDroid, which allows taint tracking and analysis to track privacy leaks in Android.
Their prototype is based on a modified version of the Dalvik virtual machine which runs Android applications.
Similarly, Costa et al. \cite{Costa2010} extends the Java virtual machine for mobile devices (Java ME) for adding runtime monitoring capabilities.
On the contrary, our feasibility study indicates that it is possible to achieve runtime monitoring in an unmodified Android system.

Recently, Burgera et al. \cite{Burguera2011} presented an approach to detect malware based on collected operating system calls.
Runtime monitoring can be done at different granularity levels. 
While the approach described by Burgera et al. is at the OS call level, we aim at providing runtime monitoring at the API call level, i.e. much more 
fine-grained and closer to the application domain of mobile applications.

Davis et al. \cite{Davis:2012:IARMDroid} presented an Android Application rewriting framework prototype, and discussed its use for monitoring an 
application, and for implementing fine-grained Access Control. 

Finally, Shabtai et al. detects malware based on the collection and analysis of various system metrics, such as CPU usage, number of packets sent through the Wi-Fi, etc.
This is an indirect way of detecting malware behavior. Again, by monitoring API calls, we observe the application behavior directly. 
The empirical results presented in this paper shows that this is actually possible.

\paragraph{\bf Advertisement Permissions Separation}
Shekhar et al. \cite{adsplit} proposed a new Android advertisement system 
that would allow to have an application and its advertisement module 
to run in different processes, and hence have a different permission set. 
This new system has to be manually inserted into the application during 
the development phase, since no automated application modification is 
provided. 

Pearce et al. \cite{addroid} made the case for an advertisement framework 
that would be integrated inside the Android platform. Every developer 
would be able to use the custom-built API that would be available 
on Android devices. This approach requires a modification of the Android 
framework, and that a given user has a device with a Android version 
embeding this advertisement system.

\paragraph{\bf Permission Policy}

Erlingsson et al. and Evans et al. \cite{erlingsson2003inlined, irm-00, naccio} were the first to manipulate bytecode to weave a security policy directly in a Java programs.
Their Inline Reference Monitor (IRM) technique allows (1) to completely separate the program development from the policy definition and (2) to have a policy mechanism independent of the Java Virtual Machine on which the program is running.
We also weave the security policy directly in Android applications, obtaining robust Android applications whose security policy is independent of the Android system on which they are running. 
 
Closest to our work are two Dalvik bytecode manipulation systems: I-Arm Droid \cite{davis2012arm} and Mr. Hide \cite{jeon2011dr}. 
The main difference is that our approach runs in-vivo whereas theirs does not.

In-vivo bytecode manipulation is also achieved by AppGuard \cite{backes2012appguard, backes2013appguard}.
However, the approach is based on \emph{dexlib} a bytecode manipulation library which does not offer an abstract representation like Jimple with Soot. 
Thus, more advanced reasoning on the bytecode (on graphs for instance) is difficult with their approach.

Redirecting methods of interest to a monitor is the basic of IRM.
Von Styp-Rekowsky et al. present a novel approach by modifying the equivalent of Dalvik function pointers at runtime \cite{von2013idea}.
Such an approach reduces the overhead and could easily be adopted by our fine-grained permission system.

Xu et al. present Aurasium \cite{xu2012aurasium}, another permission management system.
It does operate at the level of C libraries and redirect low level functions of interest to the monitor.
Operating at this low level makes it difficult to inject fake values and to differentiate between normal and Java-level security relevant operations.

Reddy et al. \cite{reddy2011application} claim that security of 
the Android platform would be improved by creating 
``application-centric permissions'' i.e. permissions expressing 
what an application can do rather than current Android permissions 
that express what resource an application can use. They wrote a 
library that allows the `application-centric permissions'' to be 
managed. In addition, they started developing a tool called ``redexer'' 
whose aim is to automatically rewrite existing applications in order for 
them to use these new permissions. 

Nauman et al. \cite{Nauman2010} extended the Android policy-based 
security model so that it can enforce constraints at runtime. The tool 
they created, called Apex, allows a user to express limits imposed to an 
application's use of any permission: For example, it becomes possible 
with Apex to grant the \texttt{SEND\_SMS} permission to any given 
application while ensuring that this application will not be able to 
send more than a user-defined amount a text message each day. The user 
also has the possibility to change her mind, and to totally prevent 
the application from sending short messages; This is an important 
improvement over the stock Android OS because it allows users to 
specify a much finer-grained policy, instead of having to choose between 
either granting an application every permission it may request at 
installation time or not installing this application. However, this 
approach requires modifications deep inside the Android framework, 
and hence would need to be backed by Google and integrated into future 
versions of Android if it was to be widely used.

\section{Conclusion}
\label{sec:conclusion}

The toolchain we propose and evaluate in this paper is a milestone that respond to the recent claim of  Stravou et al. \cite{Computer2012_Stavrou} about the urgent need for bytecode analysis to perform in-vivo  security checks on mobile phones.
We have 1) proposed a tool chain allowing the manipulation, instrumentation and analysis 
of Android bytecode and 2) shown that it is possible to run the tool chain in a reasonable amount of time 
directly on unmodified smartphones with unmodified Android software stack.
Concretely, our experiment shows that with ASM, 39 (30\%) applications of our 
dataset can be instrumented in less than 952 seconds (with a median time of 120s).
Moreover, we discuss specific limitations that we observed, such as the hard-coded heap size of Android systems. 

We believe that those various limitations could be quickly overcome, at least for two main reasons. 
First, we used off-the-shelf Java tools that are not optimized to run on environments where resources (memory/CPU) are limited, and there may be possibilities of significant optimization. 
Second, the hardware and OS evolution of smartphones will make it possible to process ever bigger Android applications (for instance, on Android 4, the default size of the heap is twice as large as in the previous version).  

We are currently working on other use cases. 
In particular, we are implementing a behavioral malware detection approach that is set up and run on the smartphone. This approach involves instrumenting the bytecode to redirect API method calls to stubs responsible for detecting malicious behavior.

%


\printbibliography



\end{document}